# Gap solitons supported by optical lattices in photorefractive crystals with asymmetric nonlocality

Zhiyong Xu, Yaroslav V. Kartashov, and Lluis Torner

*ICFO-Institut de Ciencies Fotoniques, and Universitat Politecnica de Catalunya,*

*Mediterranean Technology Park, 08860 Castelldefels (Barcelona), Spain*

We address the impact of asymmetric nonlocal diffusion nonlinearity on the properties of gap solitons supported by photorefractive crystal with an imprinted optical lattice. We reveal how the asymmetric nonlocal response alters the domains of existence and the stability of solitons originating from different gaps. We find that in such media gap solitons cease to exist above a threshold of the nonlocality degree. We discuss how the interplay between nonlocality and lattice strength modifies the gap soliton mobility.

Self-action of light in periodic nonlinear structures generates rich optical phenomena [1]. In particular, such nonlinear structures, or optical lattices, support various types of solitons that appear as defect nonlinear modes residing in gaps of Floquet-Bloch lattice spectrum. Finite gaps give rise to solitons forming due to nonlinear coupling between waves having equal longitudinal wave vector components and opposite wave vector components in the transverse direction, when both of them experience Bragg scattering [2]. Gap solitons were studied in different materials, including photonic crystals and layered microstructures [3,4], fiber Bragg gratings [5], Bose-Einstein condensates [6-8], waveguide arrays [9-11], and optically induced lattices [12,13].

Nonlocality of nonlinear response may drastically modify conditions necessary for gap soliton existence. Nonlocality strongly affects solitons even for symmetric nonlinear responses [14]. Asymmetric nonlocality exhibited by photorefractive crystals [15-18] may have an even stronger impact on the properties of solitons emerging from finite gaps, a feature which is the focus of Letter. In particular, we reveal that gap solitons in photorefractive crystals with nonlocal diffusion nonlinearity have largely asymmetric oscillating profiles and that they cease to exist when nonlocality exceeds a threshold. We also study the impact of nonlocality on mobility of gap solitons, studied previously only in focusing nonlocal media [19,20].

We consider propagation of light along the $\xi$ axis of a biased photorefractive crystal that exhibits defocusing drift and asymmetric diffusion components of the nonlinear response, in the presence of an imprinted optical lattice. The propagation dynamics is described by the nonlinear Schrödinger equation for the dimensionless complex light field amplitude $q$:

$$i\frac{\partial q}{\partial \xi} = -\frac{1}{2}\frac{\partial^2 q}{\partial \eta^2} + q|q|^2 + \mu q \frac{\partial}{\partial \eta}|q|^2 - pR(\eta)q. \qquad (1)$$

Here $\eta$ and $\xi$ stand for the transverse and longitudinal coordinates scaled to the beam width and diffraction length, respectively; the parameter $p$ characterizes the lattice depth; the function $R(\eta) = \cos(2\pi\eta/T)$ describes lattice profile, where $T$ is the modulation period, and the parameter $\mu$ stands for the strength of the nonlocal diffusion nonlinearity. The Eq. (1) can be derived from the Kukhtarev-Vinetskii material equations [15-18], and it describes propagation of light at low light intensities $I \ll I_{\text{bg}}$, where $I_{\text{bg}}$ is the intensity of background illumination. At higher intensity saturation of photorefractive nonlinear response becomes significant, but in this regime self-bending effects arising from asymmetric response dominate over lattice effects, a situation that corresponds to the opposite limit than the one addressed here. We verified numerically that for the power levels where nonlocal and lattice effects compete on similar footing, full saturable model and Eq. (1) give qualitatively similar results. The strength of the diffusion component of nonlinear response in Eq. (1) increases with decrease of beam width. When $\mu \to 0$ (for broad beams) one recovers the local medium. The nonlocality is significant for tightly focused light beams, however in most of physically realistic situations $\mu < 1$. Equation (1) is derived taking into account the lowest-order space-charge field effects [17], an approach that is justified for most of photorefractive crystals with paraxial light beams. The largest correction to Eq. (1) arising on account of higher-order space-charge field effects results in a correction to the parameter $\mu$, while corrections proportional to $\partial^2|q|^2/\partial\eta^2$ and $(\partial|q|^2/\partial\eta)^2$ are negligible for paraxial beams (e.g., for a beam width of 10 $\mu$m these terms are of the order of $10^{-3}$). Equation (1) conserves the total energy flow $U = \int_{-\infty}^{\infty} |q|^2 \, d\eta$.

To study the conditions necessary for existence of gap solitons we analyze the Floquet-Bloch lattice spectrum by solving linearized Eq. (1) with $q(\eta, \xi) = w(\eta) \exp(ib\xi + ik\eta)$, where $b$ is the propagation constant, $k$ is a Bloch wave number, and $w(\eta) = w(\eta + T)$. A typical spectrum $b(p)$ is depicted in Fig. 1(a) for $T = \pi/2$. All possible propagation constant values are arranged into bands, where Eq. (1) admits Bloch wave solutions. These bands are separated by gaps where periodic waves do not exist. The Floquet-Bloch spectrum possesses a single semi-infinite gap and infinite number of finite gaps. Solitons emerge as defect modes whose propagation constants are located inside gaps. Lowest-order (odd) solitons existing in semi-infinite gap in focusing media, rely on mechanism of total internal reflection and feature bell-shaped profiles [2]. Solitons from finite gaps rely on mechanism of Bragg reflection and feature transverse shape oscillations [11].

Stationary solutions of Eq. (1) were found in the form $q(\eta, \xi) = w(\eta) \exp(ib\xi)$, where $w$ is a real function. To analyze stability we looked for perturbed solutions $q = (w + u + iv) \exp(ib\xi)$, where the real $u(\eta, \xi)$ and imaginary $v(\eta, \xi)$ perturbation parts can grow with a complex rate $\delta$ upon propagation. Linearization of Eq. (1) around $w(\eta)$ yields the eigenvalue problem

$$\delta u = -\frac{1}{2}\frac{d^2 v}{d\eta^2} + bv - pRv + 2\mu w \frac{dw}{d\eta} v + w^2 v,$$
$$\delta v = \frac{1}{2}\frac{d^2 u}{d\eta^2} - bu + pRu - 4\mu w \frac{dw}{d\eta} u - 3w^2 u - 2\mu w^2 \frac{du}{d\eta},$$
(2)

that we solved numerically. At $\mu \ne 0$ gap solitons feature strongly asymmetric shapes due to the action of diffusion nonlinearity (Fig. 2). For a fixed $b$ the position of integral gap soliton center shifts in positive direction of $\eta$-axis with increase of $\mu$. Energy flow is a nonmonotonic function of propagation constant (Figs. 1(b), 1(f)). It decreases with $b$ in most part of existence domain, except for the very narrow region near upper cutoff on $b$ where $dU/db > 0$. There exist also a lower cutoff for soliton existence. In clear contrast to local case ($\mu = 0$) solitons do not occupy the whole gap at $\mu > 0$, i.e. cutoffs typically do not coincide with gap edges (see Fig. 1(a) where domain of gap soliton existence at $\mu > 0$ is mapped onto band-gap spectrum). Physically, lower (high-power) cutoff appears due to the interplay between nonlocality and lattice strengths. The impact of nonlocality causing soliton profile deformation rapidly increases with peak amplitude, so that at certain $\mu$ lattice cannot prevent high-energy solitons from bending. Thus, lower cutoff start increasing with $\mu$ when it exceeds a threshold (Figs. 1(c),1(e)). Contrary to expectations that nonlocality affects strongly only solitons with high peak intensity, we found that upper (low-power) cutoff decreases with $\mu$. This indicates that nonlocality breaks energy exchange balance between waves with opposite transverse wave vectors resulting in formation of gap solitons when amplitudes of these waves become too small. At fixed $p$ soliton existence domain shrinks with increase of $\mu$, so that solitons cease to exist when $\mu$ exceeds a critical value (Figs. 1(c), 1(e)). The critical value increases with the lattice depth. For fixed $\mu$ gap solitons may exist only above a minimal lattice depth, that is much higher for second-gap solitons than for first-gap solitons (Fig. 1(a)). The strongest localization of gap solitons is achieved deep inside existence domains, while near cutoffs solitons become spatially extended and strongly asymmetric. Linear stability analysis indicates that first-band solitons are stable in most part of their existence domain. At $\mu \ll 1$ we found regions of weak oscillatory instability near the lower and upper cutoffs, while at moderate values of $\mu$ we encountered only a narrow domain of exponential instability near the upper cutoff, where $dU/db > 0$ (Fig. 1(d)). Asymmetric nonlocality destabilizes second-band solitons. Direct simulations of Eq. (1) in the presence of broadband input noise confirmed the above results in all cases. Namely, first-band solitons remain stable and propagate undistorted (Fig. 3(a)).

Importantly, we found that mobility of gap solitons with low and even moderate amplitudes covering several lattice periods is substantially enhanced at $\mu > 0$. Strong enough asymmetric diffusion response may cause a significant drift of gap solitons. Illustrative examples are shown in Fig. 3(b), where we launched gap soliton obtained for $\mu = 0$ into medium with $\mu > 0$ without any input tilt (approximations to such solitons can be generated experimentally, e.g. with spatial light modulators). Notice the low level of radiation losses that accompany the excitation and propagation of such soliton. We found that curvature of propagation trajectory increases with increase of nonlocality degree (see Fig. 4(a)), while the trajectory itself is close to a parabolic one at the initial stage of propagation and almost linear for large propagation distances. Changing the lattice depth for fixed $\mu$ also enables control of gap soliton mobility (Fig. 4(b)). While strong enough lattices support immobile solitons, in shallow lattices solitons drift with a curvature that increases with decrease of lattice strength.

Another practically important issue is the excitation of well-localized gap solitons covering only a few lattice periods. We found that they can be excited with a single Gaussian beam $A\exp(-\eta^2/\eta_0^2)$ with properly selected width and amplitude. Figures 3(c) and 3(d) illustrate the typical excitation dynamics. The peak of the excited well-localized gap soliton may be located away of the input channel. Under the action of asymmetric response the soliton center may jump into a neighboring lattice channel, with the number of jumps depending on the amplitude of input beam. The implications for soliton control and routing are readily apparent.

# References with titles

# References without titles

**Figure captions**

Figure 1.   (a) Band-gap structure of periodic lattice and domains of existence of gap solitons in the presence of nonlocal nonlinearity with $\mu = 0.4$. Bands are shaded and gaps are shown white. (b) Energy flow vs propagation constant for soliton from the first finite gap at $p = 3$ and $\mu = 0.5$. Points marked by circles correspond to profiles shown in Figs. 2(a) and 2(b). (c) Domains of existence of solitons from first finite gap on $(\mu,b)$ plane. (d) Real part of perturbation growth rate vs propagation constant for soliton from first finite gap at $p = 3$ and $\mu = 0.1$. (e) Domains of existence of solitons from second finite gap on $(\mu,b)$ plane. (f) Energy flow vs propagation constant for soliton originating from second finite gap at $p = 10$ and $\mu = 0.05$. Point marked by circle corresponds to profile shown in Fig. 2(c).

Figure 2.   Profiles of solitons from the first finite gap, with $b = -1.2$ (a) and $b = -2.9$ (b) at $p = 3$ and $\mu = 0.5$. Profiles of solitons from the second finite gap, when $\mu = 0.05$ (c) and $\mu = 0.28$ (d) at $b = -9$ and $p = 10$. Gray regions correspond to $R(\eta) \leq 0$ and in white regions $R(\eta) > 0$.

Figure 3.   (a) Stable propagation of solitons from first finite gap corresponding to $b = -0.57$, $p = 3$, and $\mu = 0.1$ in the presence of white input noise with variance $\sigma_{\text{noise}}^2 = 0.01$. (b) Drift of soliton from first finite gap corresponding to $b = -0.6$, $p = 3$, and $\mu = 0$, launched into nonlocal medium with $\mu = 0.07$. Excitation of gap solitons by Gaussian beam with amplitude $A = 1.6$ (c) and 1.8 (d) at $p = 2$ and $\mu = 0.2$.

Figure 4.   Trajectories of propagation for soliton from first finite gap launched into nonlocal medium. In (a) we set $b = -0.6$, $p = 3$, and vary nonlocality degree. In (b) we set $b = -1$, $\mu = 0.2$, and vary lattice depth.

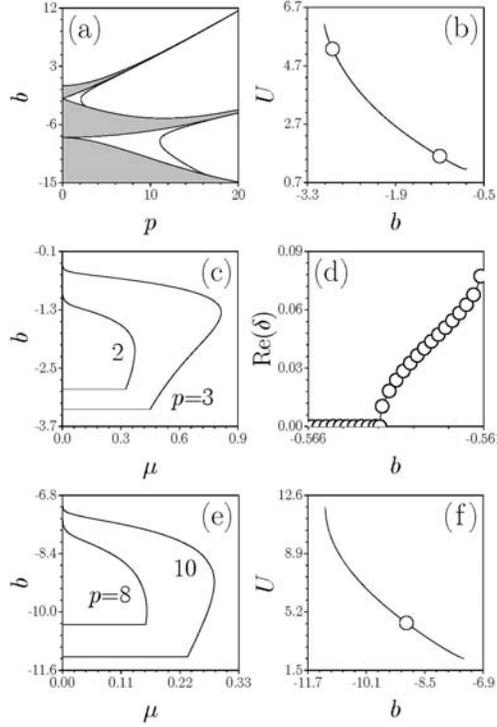

Figure 1. (a) Band-gap structure of periodic lattice and domains of existence of gap solitons in the presence of nonlocal nonlinearity with $\mu = 0.4$. Bands are shaded and gaps are shown white. (b) Energy flow vs propagation constant for soliton from the first finite gap at $p = 3$ and $\mu = 0.5$. Points marked by circles correspond to profiles shown in Figs. 2(a) and 2(b). (c) Domains of existence of solitons from first finite gap on $(\mu,b)$ plane. (d) Real part of perturbation growth rate vs propagation constant for soliton from first finite gap at $p = 3$ and $\mu = 0.1$. (e) Domains of existence of solitons from second finite gap on $(\mu,b)$ plane. (f) Energy flow vs propagation constant for soliton originating from second finite gap at $p = 10$ and $\mu = 0.05$. Point marked by circle corresponds to profile shown in Fig. 2(c).

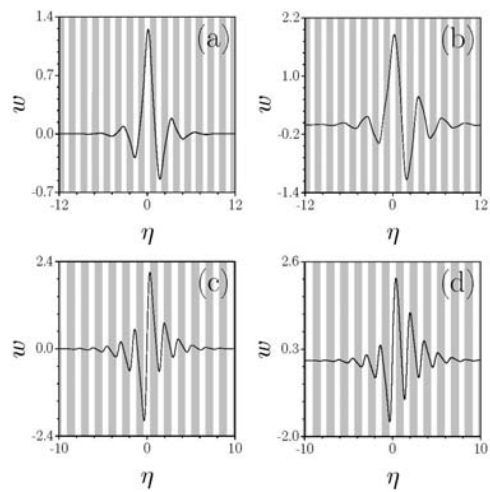

Figure 2. Profiles of solitons from the first finite gap, with $b = -1.2$ (a) and $b = -2.9$ (b) at $p = 3$ and $\mu = 0.5$. Profiles of solitons from the second finite gap, when $\mu = 0.05$ (c) and $\mu = 0.28$ (d) at $b = -9$ and $p = 10$. Gray regions correspond to $R(\eta) \leq 0$ and in white regions $R(\eta) > 0$.

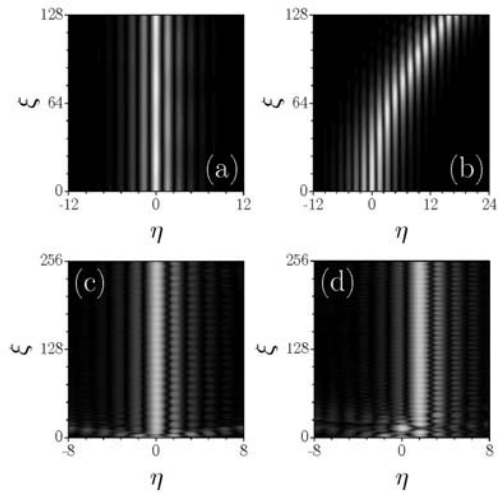

Figure 3. (a) Stable propagation of solitons from first finite gap corresponding to $b = -0.57$, $p = 3$, and $\mu = 0.1$ in the presence of white input noise with variance $\sigma^2_{\text{noise}} = 0.01$. (b) Drift of soliton from first finite gap corresponding to $b = -0.6$, $p = 3$, and $\mu = 0$, launched into nonlocal medium with $\mu = 0.07$. Excitation of gap solitons by Gaussian beam with amplitude $A = 1.6$ (c) and 1.8 (d) at $p = 2$ and $\mu = 0.2$.

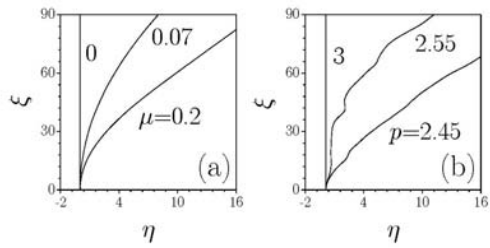

Figure 4. Trajectories of propagation for soliton from first finite gap launched into nonlocal medium. In (a) we set $b = -0.6$, $p = 3$, and vary nonlocality degree. In (b) we set $b = -1$, $\mu = 0.2$, and vary lattice depth.